\documentclass[aps,prl,twocolumn,superscriptaddress]{revtex4} 
\usepackage[T1]{fontenc}
\usepackage{amssymb}

\usepackage{hyperref}
\hypersetup{
    bookmarks=true,         
    unicode=true,          
    pdftoolbar=false,        
    pdfmenubar=true,        
    pdffitwindow=true,      
    pdftitle={My title},    
    pdfauthor={Author},     
    pdfsubject={Subject},   
    pdfnewwindow=true,      
    pdfkeywords={keywords}, 
    colorlinks=true,       
    linkcolor=red,          
    citecolor=black,        
    filecolor=magenta,      
    urlcolor=cyan           
}

\usepackage[all]{hypcap}
\usepackage{graphicx, subfigure}
\usepackage{amsmath}
\usepackage{natbib}
\usepackage{layouts}
\usepackage{threeparttable}
\usepackage{footnote}
\usepackage{booktabs}
\usepackage{scrextend}
\usepackage{hyperref}
\usepackage{bm}
\usepackage[toc]{appendix}
\usepackage{color}

\begin{document}

\title{Dynamical stability and superconductivity of (Ce,La)H$_9$ under high pressure}

\author{Prutthipong Tsuppayakorn-aek} 
\email{prutthipong.tsup@gmail.com}
\affiliation{Extreme Condition Physics Research Laboratory and Center of Excellence in Physics of Energy Materials (CE:PEM), Department of Physics, Faculty of Science, Chulalongkorn University, Bangkok 10330, Thailand}
\affiliation{Thailand Center of Excellence in Physics , Ministry of Higher Education, Science, Research and Innovation, 328 Si Ayutthaya Road, Bangkok 10400, Thailand}

\author{Udomsilp Pinsook} 
\affiliation{Department of Physics, Faculty of Science, Chulalongkorn University, 10330 Bangkok, Thailand}

\author{Thiti Bovornratanaraks} 
\affiliation{Extreme Condition Physics Research Laboratory and Center of Excellence in Physics of Energy Materials (CE:PEM), Department of Physics, Faculty of Science, Chulalongkorn University, Bangkok 10330, Thailand}
\affiliation{Thailand Center of Excellence in Physics , Ministry of Higher Education, Science, Research and Innovation, 328 Si Ayutthaya Road, Bangkok 10400, Thailand}

\date{\today}

\begin{abstract} 
\indent{Recent experiments have shown that CeH$_{9}$ and (Ce,La)H$_{9}$ can be synthesized under high pressure between 90-170GPa and become a superconductor with a high value of superconducting critical temperature ($T_c$) between 100-200K. In this work, we performed a theoretical study of a (Ce,La)H$_{9}$ compound where the Ce:La ratio is equal to 1. We used the density functional theory and the {\it ab initio} molecular dynamics (AIMD) method. From the phonon dispersion, there exist some unstable modes around the K-point phonons. Then, we performed AIMD simulation at around 203K and found that the compound becomes stable. The superconducting spectral function can be calculated. We found that $\lambda$ is as high as $3.0$ at 200GPa. By using Allen-Dynes-modified McMillan equation in the strong coupling regime, we found that $T_c = 87K$ at 200GPa.}

\end{abstract}

\maketitle 

\section{Introduction}
In 1968, Ashcroft suggested that metallic hydrogen can become a conventional superconductor with a high value of critical temperature [1]. He also pointed out that metal hydrides could become a superconductor [2,3]. Since then, the search for metallic hydrogen and hydride superconductors have been extensively explored by both theory and experiments [4-20]. Metal superhydrides have recently attracted much more of the scientific attention because they can be synthesized and characterized under high pressure. Furthermore, some of them have been found to be a superconductor with a close-to-room-temperature superconducting critical temperature ($T_c$). For instance, the T$_{c}$ of LaH$_{10}$ has been measured to be 250K under high pressure [21].

In 2019, cerium superhydride CeH$_{9}$ has been synthesized under high pressure between 80 and 100GPa by using the laser-heated diamond anvil cell combined with synchrotron X-ray diffraction [14]. It was found that CeH$_{9}$ is a hexagonal structure with a space group of P6$_{3}$/mmc. Its $T_{c}$ has been theoretically estimated to be 105 - 117K [14]. Recently, the (La,Ce)H$_x$ compounds have been successfully synthesized at pressure lower than 130GPa by Chen, et al. [22]. They discovered that the compound resembles (La,Ce)H$_9$ with $T_{c}$ equal to 176K at 100GPa. 

At the current stage, the trend has moved from searching for the superconductivity of a binary compound to searching for the superconductivity of a ternary hydride compound. This is because the mixing of different atomic species in the making of ternary compounds offers a slight and perhaps controllable modification to the electronic and dynamical properties of the hydride compounds, which sometimes tends to enhance $T_c$ [12, 17, 20, 23-26]. For example, Mg can be used as a substitution for Ca in the CaH$_{6}$ compound [17]. It is found that Mg$_{0.5}$Ca$_{0.5}$H$_{6}$ compound is stable at 200GPa and become a superconductor with $T_c \approx$ 288K. 

In this work, we performed the simulation of the Ce/La substituted of (Ce,La)H$_{9}$ at 200GPa, as both CeH$_9$ and LaH$_9$ share a common hexagonal structure with a space group of P6$_{3}$/mmc [14, 27, 28]. The hexagonal structure is also suggested by the experiment for the La-Ce-H compounds [22]. We calculated the electron-phonon interaction and evaluated the Eliasberg spectral function by the density functional theory (DFT). The $T_c$ can then be approximated. We also preformed the {\it ab initio} molecular dynamics (AIMD) to examine the dynamical stability of (Ce,La)H$_{9}$.

\section{Computational details}
In this work, we performed the DFT calculations with the generalized gradient approximation of the Perdew--Burke--Ernzerhof (GGA-PBE) functional [29] for the exchange-correlation functional, and the projector augmented wave (PAW) method [30], as implemented in the quantum espresso (QE) [31,32]. For the calculation of the electron-phonon interaction, a plane-wave energy cutoff of 60Ry was used. The electron-phonon coupling (EPC) matrix elements were computed in the first Brillouin-zone (BZ) on 4$\times$4$\times$2 q-meshes using individual EPC matrices obtained with 24$\times$24$\times$16 k--point meshes. The Allen-Dynes-modified McMillan equation [33] was used for the estimating T$_{c}$.

For the electronic structure, we used a plane-wave basis set with the cutoff energy of 600eV and the first BZ with 12$\times$12$\times$8 k-point meshes. The pseudocore radii of Ce, La, and H are 2.57Bohr, 2.80Bohr, and 1.1Bohr, which are small enough to ensure that no core overlapping will occur at 200GPa.  For {\it ab initio} molecular dynamics (AIMD), the simulation was performed by using the $NPT$ ensemble with 160 atoms per supercell. The AIMD simulation was carried out at 203K and 200GPa. Both the electronic structure and AIMD were calculated by using the Vienna ab initio simulation package (VASP) [34]. 


\section{Results and discussion}
\begin{figure}[ht]
\includegraphics[width=0.4\textwidth]{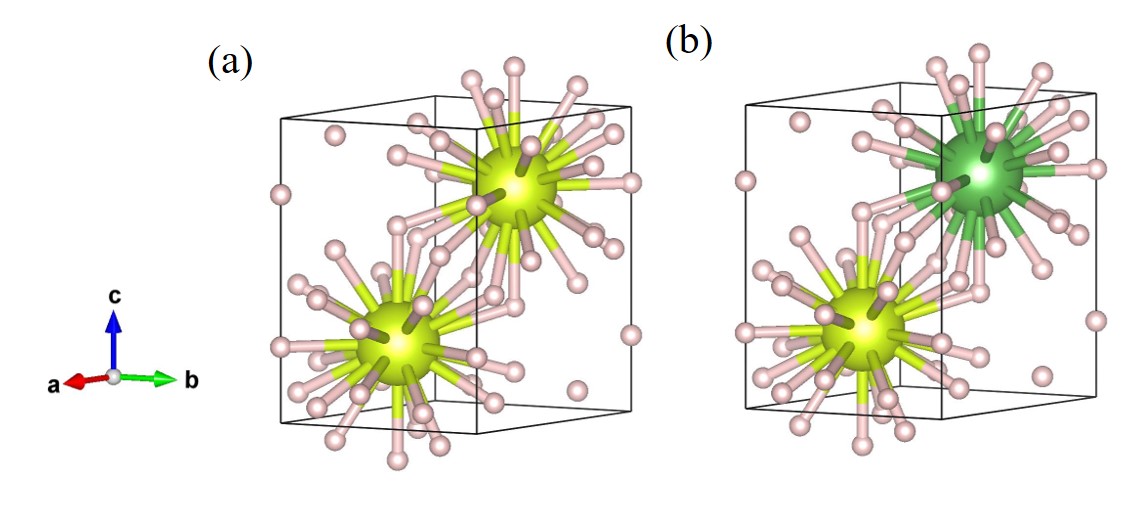}		
\caption {(a) The hexagonal structure of CeH$_{9}$, where the yellow spheres represent the Ce atoms and the pink spheres represent the H atoms. (b) The hexagonal structure of (Ce,La)H$_{9}$ , where the yellow sphere represents the Ce atom, the green sphere represents the La atom, and the pink spheres represent the H atoms.}
\label{fig1}
\end{figure}

As both CeH$_9$ and LaH$_9$ share a common hexagonal structure with a space group of P6$_{3}$/mmc [14, 27, 28], as shown in Fig.~\ref{fig1}(a). The unit cell of the P6$_{3}$/mmc structure contains two metal sites. We modelled the (Ce,La)H$_9$ compound by inserting an La atom into one of the metal site, and a Ce atom into the others. The Ce:La ratio is equal to 1. The symmetry is reduced from the space group of P6$_{3}$/mmc to P${\bar 6}$m2, as shown in Fig.~\ref{fig1}(b). 

\begin{figure}[ht]
\includegraphics[width=0.5\textwidth]{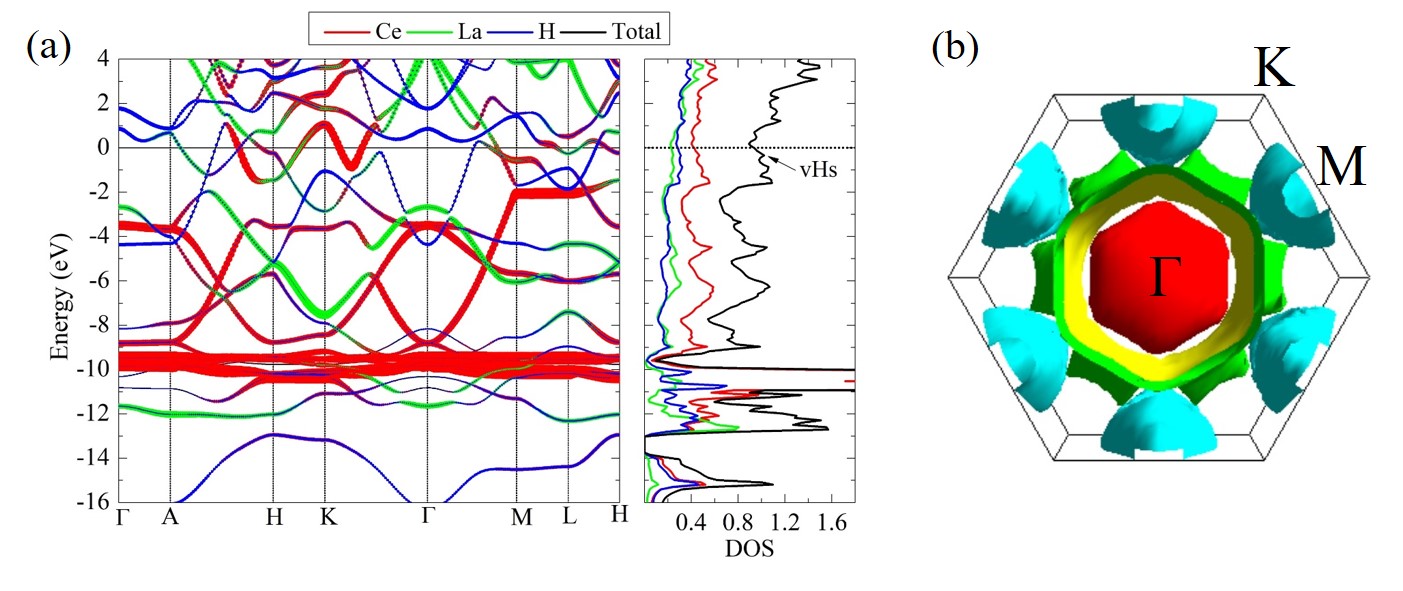}			
\caption {Electronic structures of (Ce,La)H$_{9}$  at 200 GPa. (a) The band dispersions (left) and the projected density of states (PDOS). (b) The Fermi surface topology (FST)}
\label{fig2}
\end{figure}

The electronic structures are shown in Fig.~\ref{fig2}(a). Under high pressure, the (Ce,La)H$_{9}$  compound becomes metallic. The band dispersions exhibit the weaving of the up-running bands from below E$_{F}$ and down-running bands above E$_{F}$, similar to those of CeH$_{9}$ [14], CeH$_{10}$ [35], and LaH$_{10}$ [36]. The corresponding projected density of states (PDOS) of (Ce,La)H$_{9}$  is also shown in Fig.~\ref{fig2}(a). There are several places near the Fermi level where the electronic structures exhibit flat dispersions, such as around the H-point and M-point. These flat dispersions lead to the so-called van Hove singularity (vHs). The vHs is marked by an arrow in Fig.~\ref{fig2}(a). The Fermi surface topology (FST) is shown in Fig.~\ref{fig2}(b). The FST exhibits a number of parallel surfaces, which should lead to Fermi surface nesting and enhance superconductivity.  

As T$_{c}$ can be related to the number of electronic states at the Fermi surface, it is useful to consider the local density of states (LDOS) [37,38], which is defined as 
\begin{equation}\label{eq:LDOS}
N(E,\mathbf{r}) = \sum_{n}\int_{}^{}\frac{d^3k}{(2\pi)^3}\delta(E-\epsilon_{nk})|\psi_{nk}(\mathbf{r})|^2,
\end{equation}
where $\psi$($\mathbf{k}$)($\mathbf{r}$) and $\epsilon$$_{nk}$ are the Kohn-Sham eigenfunctions and eigenvalues of the system. The LDOS allows us to evaluate the number of the electronic states at the Fermi level, which can be related to electron pairings in the superconducting state. The LDOS indicates that the Fermi surface of the (Ce,La)H$_{9}$  compound contains 0.91 states/eV. 

\begin{figure}[ht]
\includegraphics[width=0.5\textwidth]{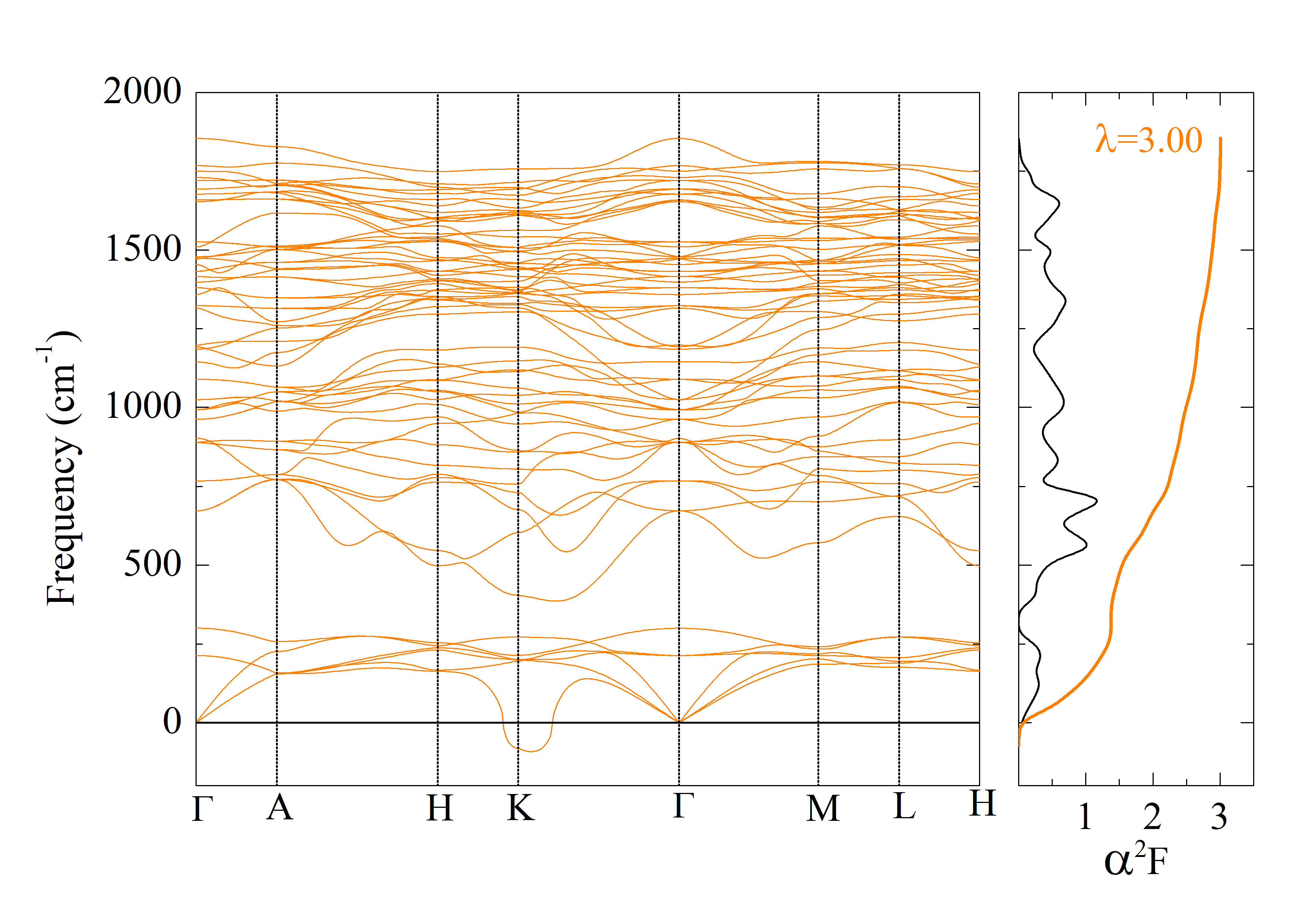}		
\caption {(Left) Phonon dispersion curves of (Ce,La)H$_{9}$  at 200GPa. (Right) The Eliashberg spectral function $\alpha$$^{2}$$F(\omega)$ and the corresponding accumulated $\lambda(\omega)$ at 200GPa.}
\label{fig3}
\end{figure}

Next, we calculated the isotropic Eliashberg spectral function, which is written as
\begin{equation}\label{eq:Eliashberg spectral function}
\alpha^{2}F(\omega) = \frac{1}{N_F} \sum_{\nu,k,k'}|g^{\nu}_{kk'}|^2\delta(\epsilon_k)\delta(\epsilon_k')\delta(\omega-\omega_{q\nu}),
\end{equation}
where N$_{F}$ is the density of states at the Fermi level, k and k$'$ are the electronic states, and q are the phonon modes. N$_{k}$ (N$_{q}$) is the total number of k (q) points in the simulation cell, $\epsilon$$_{k}$ is the energy eigenvalue of the Kohn-Sham state with respect to the Fermi level, and g$^{\nu}_{kk'}$ is the electron-phonon matrix elements for the scattering between the electronic states k and k$'$ via a phonon with wave vector $q = k'-k$, and $\omega$$_{q\nu}$ are the frequencies of phonon modes q and phonon band $\nu$. The calculated Eliashberg spectral function $\alpha$$^{2}$$F(\omega)$ is shown in the right panel of Fig.~\ref{fig3}. 

The corresponding phonon dispersions are also shown in the left panel of Fig.~\ref{fig3}. It is obvious that the acoustic modes are well-seperated from the optical modes. The optical modes are in the frequency range between 360cm$^{-1}$ and 1814cm$^{-1}$. The acoustic modes are in the frequency range between 0-281cm$^{-1}$, and contain several imaginary frequencies around the K-point phonon modes, as shown in the left panel of Fig.~\ref{fig3}. These modes indicate that the (Ce,La)H$_{9}$  compound with the hexagonal structure would indeed be an unstable structure. We will reexamine this point in more detail later.  
 
From $\alpha$$^{2}$$F(\omega)$, we can evaluate the average coupling strength $\lambda$ as a function of $\omega$ by 
\begin{equation}\label{eq:lambda}
\lambda(\omega) = 2 \int_{0}^{\omega}d\omega'\frac{\alpha^{2}F(\omega')}{\omega'},
\end{equation}
This is the so-called accumulated $\lambda(\omega)$. The result of $\lambda(\omega)$ is also shown in the right panel of Fig.~\ref{fig3}. The total $\lambda$ can be obtained by integrating over all available frequencies. We found that for (Ce,La)H$_{9}$, $\lambda$ is as high as 3.0. Thus, the electron-phonon interaction in (Ce,La)H$_{9}$ enters the strong coupling regime. As a consequence, we chose the Allen-Dynes-modified McMillan equation with strong coupling correction for describing the superconducting critical temperature, as follows;
\begin{equation}\label{eq:Allen-Dynes}
T_{c} = f_1f_2\frac{\omega_{log}}{1.2} \exp \Big[ -\frac{1.04(1+\lambda)}{\lambda-\mu^*(1+0.62\lambda)} \Big],
\end{equation} 
We chose $\mu^* = 0.1$. We found that f$_{1}$ and f$_{2}$ are 1.22 and 1.20, respectively. The $\omega_{log}$ is 327K. We found that T$_{c}$ is 87K. Despite of the large value of $\lambda$ and N$_{F}$, the $\omega_{log}$ is very low. This is because the bandwidth of $\alpha$$^{2}$$F(\omega)$ is very large [15]. As a result, T$_{c}$ becomes not as high as we might expect. 

\begin{figure}[ht]
\includegraphics[width=0.5\textwidth]{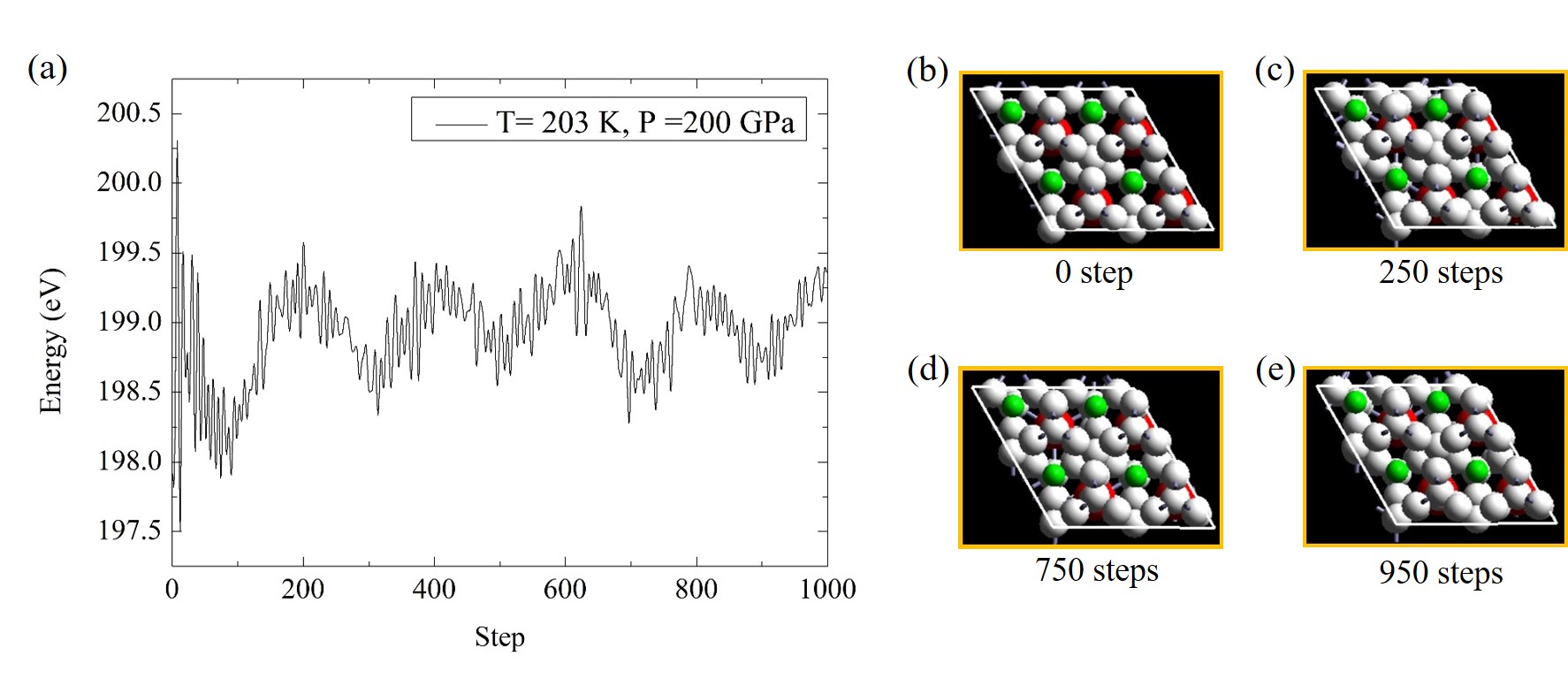}		
\caption {(a) Energy as a function of time of (Ce,La)H$_{9}$ from AIMD simulations (b-d) the snapshots of the AIMD simulations at different time steps.}
\label{fig4}
\end{figure}

At this stage, we discuss the structure stability. From the phonon dispersion in the left panel of Fig.~\ref{fig3}, there exist some unstable phonon modes at 0K. In order to take the effects of temperature into the account, we performed the AIMD simulation, with the NPT ensemble. The simulation cell was construct from a 2$\times$2$\times$2 hexagonal supercell with 160 atoms. We chose temperature of 203K and pressure of 200GPa. The energy as a function of time is shown in Fig.~\ref{fig4} (a). The time step is 1 fs. We performed the AIMD integration to 1,000 time step. After the simulation is in equilibrium, we took several snapshots, as shown in Fig.~\ref{fig4} (b)-(e). The hexagonal structure remains stable during the simulation time. The AIMD simulation suggests that the effects of temperature would help stabilize the hexagonal structure of the (Ce,La)H$_{9}$ compound. 
  
\section{Conclusion}
\indent{We studied the dynamical stability and superconductivity of the (Ce,La)H$_{9}$ compound. The symmetry is reduced from the space group of P6$_{3}$/mmc, as of CeH$_{9}$ and LaH$_{9}$, to $P{\bar 6}m2$. The phonon dispersions showed some unstable phonon modes around the K-point phonons. However, the Elaisberg spectral function can still be evaluated. We found that $\lambda$ is as high as 3.0 and $N(E_F)$ is as large as 0.91 states/eV. Despite of high values of these favorable parameters, the $\omega_{log}$ appears to be very low. This results in a low $T_{c}$ = 87K.

\section{ACKNOWLEDGMENTS}
We acknowledge the Computational Materials Physics project SLRI, Thailand for providing support for computational resources. We thank the Chalawan HPC cluster, which is operated and maintained by the National Astronomical Research Institute of Thailand (NARIT) under the Ministry of Science and Technology of the Royal Thai government. This research project was supported by the Second Century Fund (C2F), Chulalongkorn University. This project is funded by National Research Council of Thailand (NRCT): (NRCT5-RSA63001-04). This research is partially funded by Chulalongkorn University; Grant for Research.


\section{References}
\indent{[1] N. W. Ashcroft, Metallic hydrogen: A high-temperature
superconductor?, Phys. Rev. Lett. 21, 1748 (1968).}

\indent{[2] N. W. Ashcroft, Bridgman's high-pressure atomic destructibility and its growing legacy of ordered states, J. Condens. Matter Phys. 16, S945 (2004).}

\indent{[3] N. W. Ashcroft, Hydrogen dominant metallic alloys:
High temperature superconductors?, Phys. Rev. Lett. 92, 187002 (2004).}

\indent{[4] E. Zurek, R. Hoffmann, N. W. Ashcroft, A. R.
Oganov, and A. O. Lyakhov, A little bit of lithium does a lot for hydrogen, Proc. Natl. Acad. Sci. U.S.A. 106, 17640 (2009).}

\indent{[5] T. A. Strobel, A. F. Goncharov, C. T. Seagle, Z. Liu, M. Somayazulu, V. V. Struzhkin, and R. J. Hemley, High pressure study of silane to 150 GPa, Phys. Rev. B 83, 144102 (2011).}

\indent{[6] A. Shamp and E. Zurek. Superconducting high pressure phases composed of hydrogen and iodine, J. Phys. Chem, 6, 4067 (2015).}

\indent{[7] I. Errea, M. Calandra, C. J. Pickard, J.
Nelson, R. J. Needs, Y. Li, H. Liu, Y. Zhang, Ya. Ma, and F. Mauri, High pressure hydrogen sulfide from first principles: A strongly anharmonic phonon-mediated superconductor, Phys. Rev. Lett. 114, 157004 (2015).}

\indent{[8] A. P. Drozdov, M. I. Eremets, I. A. Troyan, V. Ksenofontov, and S. I. Shylin, Conventional superconductivity
at 203 kelvin at high pressures in the sulfur hydride system, Nature, 525, 73 (2015).}

\indent{[9] C. M. P\'epin, G. Geneste, A. Dewaele, M. Mezouar, and P. Loubeyre. Synthesis of FeH$_{5}$: A layered structure
with atomic hydrogen slabs, Science, 357, 382 (2017).}

\indent{[10] B. Liu, W. Cui, J. Shi, L. Zhu, J. Chen, S. Lin, R. Su, J. Ma, Ka. Yang, M. Xu, J. Hao, A. P. Durajski, J. Qi, Y. Li, and Y. Li, Effect of covalent bonding on the superconducting critical temperature of the H-S-Se system, Phys. Rev. B 98, 174101 (2018).}

\indent{[11] C. Heil, S. Cataldo, G. B. Bachelet,
and L. Boeri. Superconductivity in sodalite-like ytttrium hydride clathrates, Phys. Rev. B 99, 220502 (2019).}

\indent{[12] Y. Sun, J. Lv, Y. Xie, H. Liu, and Y. Ma, Route to a superconducting phase above room temperature in electron-doped hydride compounds under high pressure, Phys. Rev. Lett. 123, 097001 (2019).}

\indent{[13] X. Liang, A. Bergara, L. Wang, B. Wen, Z. Zhao, X. F. Zhou, J. He, G. Gao, and Y. Tian. Potential high-T$_{c}$ superconductivity in CaYH$_{12}$ under pressure, Phys. Rev. B 99, 100505 (2019).}

\indent{[14] N. P. Salke, M. M. D. Esfahani, Y. Zhang, I. A. Kruglov, J. Zhou, Y. Wang, E. Greenberg, V. B. Prakapenka, J. Liu, A. R. Oganov, and J. F. Lin, Synthesis of clathrate cerium superhydride CeH$_{9}$ at 80-100 GPa with atomic hydrogen sublattice, Nat. Commun, 10, 4453 (2019).}

\indent{[15] U. Pinsook, In search for near-room-temperature
superconducting critical temperature of metal superhydrides under high pressure: A review. J. Met. Mater. Miner, 30(2) (2020).}

\indent{[16] D. V. Semenok, A. G. Kvashnin, A. G.
Ivanova, V. Svitlyk, V. Yu. Fominski, A. V. Sadakov, O. A. Sobolevskiy, V. M. Pudalov, I. A. Troyan, and A. R. Oganov. Superconductivity at 161 K in thorium hydride ThH$_{10}$: Synthesis and properties, Mater. Today, 33, 36 (2020).}

\indent{[17] W. Sukmas, P. Tsuppayakorn-aek, U. Pinsook, and T. Bovornratanaraks, Near-room-temperature superconductivity of Mg/Ca substituted metal hexahydride under pressure. J. Alloys Compd, 849, 156434 (2020).}

\indent{[18] P. Tsuppayakorn-aek, U. Pinsook, W. Luo, R. Ahuja, and T. Bovornratanaraks, Superconductivity of superhydride CeH$_{10}$ under high pressure.Mater. Res. Express, 7, 086001 (2020).}

\indent{[19] H. Xie, Y. Yao, X. Feng, D. Duan, H. Song, Z. Zhang, S. Jiang, S. A. T. Redfern, V. Z. Kresin, C. J. Pickard, and T. Cui, Hydrogen pentagraphenelike structure stabilized by
hafnium: A high-temperature conventional superconductor, Phys. Rev. Lett. 125, 217001 (2020).}

\indent{[20] P. Tsuppayakorn-aek, P. Phansuke, P. Kaewtubtim, R. Ahuja, and T. Bovornratanaraks, Enthalpy stabilization of superconductivity in an alloying S-P-H system: First-principles cluster expansion study under high pressure. Comput. Mater. Sci, 190, 110282 (2021).}

\indent{[21] A. P. Drozdov, P. P. Kong1, V. S. Minkov, S. P. Besedin, M. A. Kuzovnikov, S. Mozaffari, L. Balicas, F. F. Balakirev, D. E. Graf, V. B. Prakapenka, E. Greenberg, D. A. Knyazev, M. Tkacz, and M. I. eremets, Superconductivity at 250 K in lanthanum hydride under high pressures. Nature, 569, 528 (2019).}

\indent{[22] W. Chen, X. Huang, D. V. Semenok, S Chen, K. Zhang, A. R. Oganov, and T. Cui, Enhancement of superconducting critical temperature realized in La-Ce-H system at moderate pressures, arXiv:2203.14353 (2022).}

\indent{[23] F. Fan, D.A. Papaconstantopoulos, M.J. Mehl, and B.M. Klein, High-temperature superconductivity at high pressures for H$_{3}$Si$_{x}$P$_{1-x}$, H$_{3}$P$_{x}$S$_{1-x}$, and H$_{3}$Cl$_{x}$S$_{1-x}$, J. Phys. Chem. Solids, 99, 105 (2016).}

\indent{[24] Y. Ge, F. Zhang, and Y. Yao, First-principles
demonstration of superconductivity at 280 K in hydrogen sulfide with low phosphorus substitution, Phys. Rev. B 93, 224513 (2016).}

\indent{[25] A. Nakanishi, T. Ishikawa, and K. Shimizu, First-principles study on superconductivity of P- and Cl-doped H$_{3}$S, J. Phys. Soc. Jpn. 87, 124711 (2018).}

\indent{[26] A. P. Durajski and R. Szcz\k{e}\'sniak. Gradual reduction of the superconducting transition temperature of H$_{3}$S by partial replacing sulfur with phosphorus, Physica C Supercond, 554, 38 (2018).}

\indent{[27] I. A. Kruglov, D. V. Semenok, H. Song, R. Szcz\k{e}\'sniak, I. A. Wrona, R. Akashi, M. M. D. Esfahani, D. Duan, T. Cui, A. G. Kvashnin, and A. R. Oganov, Superconductivity of LaH$_{10}$ and LaH$_{16}$ polyhydrides, Phys. Rev. B 101, 024508 (2020).}

\indent{[28] A. M. Shipley, M. J. Hutcheon, M. S. Johnson,
R. J. Needs, and C. J. Pickard, Stability and superconductivity of lanthanum and yttrium decahydrides, Phys. Rev. B, 101, 224511 (2020).}

\indent{[29] J. P. Perdew, K. Burke, and M. Ernzerhof.
Generalized gradient approximation made simple. Phys. Rev. Lett. 77, 3865 (1996).}

\indent{[30] P. E. Bl\"ochl. Projector augmented-wave method. Phys. Rev. B 50, 17953 (1994).}

\indent{[31] S. Baroni, S. D. Gironcoli, A. D. Corso,
and P. Giannozzi, Phonons and related crystal properties from density-functional perturbation theory. Rev. Mod. Phys, 73, 515 (2001).}

\indent{[32] P. Giannozzi, S. Baroni, N. Bonini, M. Calandra, R. Car, C. Cavazzoni, D. Ceresoli, G. L. Chiarotti, M. Cococcioni, I. Dabo, A. D. Corso, S. D. Gironcoli, S. Fabris, G. Fratesi, R. Gebauer, U. Gerstmann, C. Gougoussis, A. Kokalj, M. Lazzeri, L. Martin-Samos, N. Marzari, F. Mauri, R. Mazzarello, S. Paolini, A. Pasquarello, L. Paulatto, C. Sbraccia, S. Scandolo, G. Sclauzero, A. P. Seitsonen, A. Smogunov, P. Umari, and R. M. Wentzcovitch, Quantum espresso: a modular and open-source software project for quantum simulations of materials. J. Condens. Matter Phys, 21, 395502, (2009).}

\indent{[33] P. B. Allen and R. C. Dynes, Transition temperature of strong-coupled superconductors reanalyzed, Phys. Rev. B, 12, 905 (1975).}

\indent{[34] G. Kresse and J. Furthm\"uller, Efficient iterative schemes for ab initio total-energy calculations using a plane-wavebasis set, Phys. Rev. B 54, 11169 (1996).}

\indent{[35] L. Liu, C. Wang, S. Yi, K. W. Kim, J. Kim, and Jun-Hyung Cho. Microscopic mechanism of room-temperature superconductivity in compressed LaH$_{10}$, Phys. Rev. B 99, 140501 (2019).}

\indent{[36] C. Heil, G. B. Bachelet, and L. Boeri, Absence of superconductivity in iron polyhydrides at high pressures. Phys. Rev. B 97, 214510 (2018).}

\indent{[37] S. Cataldo, W. Linden, and L. Boeri, Phase diagram and superconductivity of calcium borohyrides at extreme pressures, Phys. Rev. B 102, 014516 (2020).}

\end{document}